# Glass Transitions and Critical Points in Orientationally Disordered Crystals and Structural Glassformers: ("Strong" Liquids are More Interesting Than We Thought)

C. Austen Angell and Mahin Hemmati

*Department of Chemistry and Biochemistry, Arizona State University, Tempe, AZ 85287-1604*

**Abstract.** When liquids are classified using $T_g$-scaled Arrhenius plots of relaxation times (or relative rates of entropy increase above $T_g$) across a "strong-fragile" spectrum of behaviors, the "strong" liquids have always appeared rather uninteresting[1,2]. Here we use updated plots of the same type for crystal phases of the "rotator" variety[3] to confirm that the same pattern of behavior exists for these simpler (center of mass ordered) systems. However, in this case we can show that the "strong" systems owe their behavior to the existence of lambda-type order-disorder transitions at higher temperatures (directly observable in the cases where observations are not interrupted by prior melting). Furthermore, the same observation can be made for other systems in which the glass transition, at which the ordering is arrested, occurs in the thermodynamic *ground* state of the system. This prompts an enquiry into the behavior of strong liquids at high temperatures. Using the case of silica itself, we again find strong evidence from extended ion dynamics simulations, for a lambda transition at high temperatures, but only if pressure is adjusted to a critical value. In this case the lambda point is identifiable as a liquid-liquid critical point of the type suggested for supercooled water. We recognize the possibility of exploring, *a postiori*, the consequences of rapid cooling of laboratory liquid $SiO_2$ from >5000K and multi-GPa pressures, using the phenomenology of damage-induced plasmas in optical fibers. The ramifications of these considerations will be explored to establish a "big picture"[2] of the relation of thermodynamic transitions to supercooled liquid phenomenology[4,5].

**Keywords:** glass transition, fragility, lambda transition, plastic crystals, liquid-liquid transition, high pressure,

## INTRODUCTION

The origin of the striking and anomalous dynamics of the typical glassforming liquid has eluded full understanding for more than a century. Despite the apparent difficulty of reaching an accord in interpretation of the wide range of behavior that can be observed, there is general agreement about the existence of a pattern of behavior, rather simple in form, that emerges when a given property, like viscosity, or relaxation time, is plotted in a "scaled Arrhenius" form, using the "standard" glass transition as the scaling parameter for temperature[6]. The "standard" glass transition temperature is taken as the onset temperature for the well-known heat capacity "jump", when the sample is upscanned at 20K/min after cooling at the same rate[2,7].

An example of this plot, which is now familiar in glass science, is given in Figure 1, and then compared with the equivalent diagram for systems that are not liquids, but instead have center-of-mass order (i.e. are crystals). Instead they have disorder in their molecular or ionic orientations[8] and the reorientation relaxation times, detected by dielectric relaxation measurements, are displayed in the same scaled Arrhenius form (after Brand et al[9]). The linear Arrhenius form typified by silica, and called "strong" liquid behavior, is quite uncommon amongst liquids, most liquids tending to cluster in the middle and "fragile" sides of the total pattern[6]. By contrast, amongst the rotator phases (called in general "plastic" crystals) the majority of cases are "strong" in character, and the "fragile" cases are rare[9]. This provides some motivation to look into what might underlie the "strong" behavior of these relaxing systems. The surprising answer then provokes the rest of the paper's ideas.

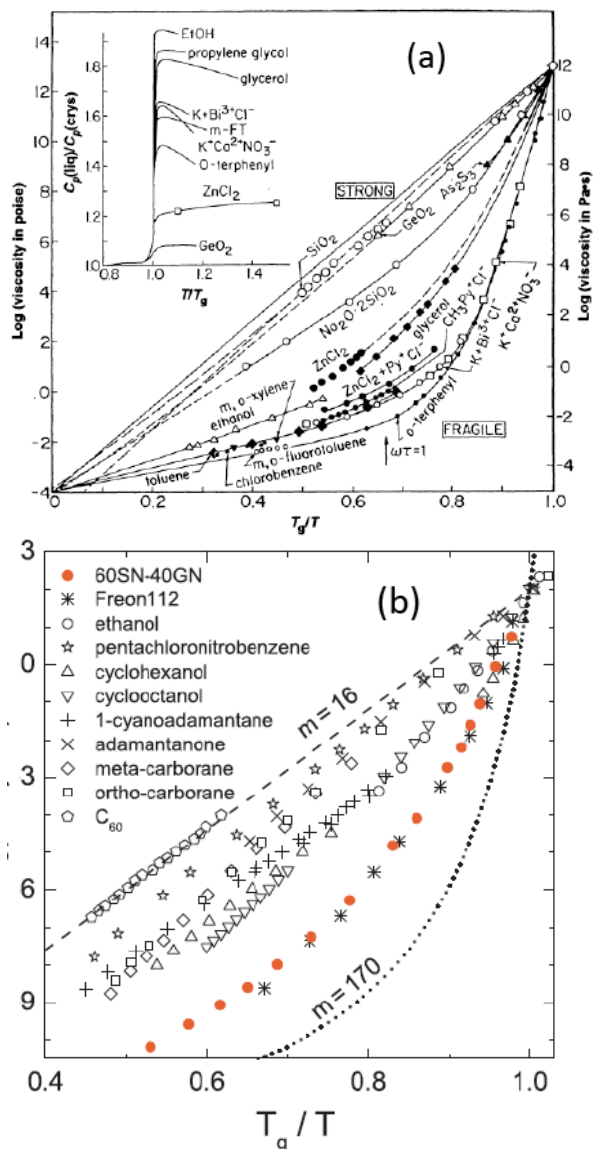

**FIGURE 1.** (a) Scaled Arrhenius plots of glassforming liquids (b) and "glassy crystal"-forming systems, showing the "strong/fragile" pattern of behavior common to each.

## WHAT THE PLASTIC CRYSTAL PHENOMENOLOGY CAN TELL US

Prominent among the plastic crystals is the case of $C_{60}$ (Buckminsterfullerite), in which the ground state packing has local dipolar character which can be excited by an electric field and so is dielectrically active[10]. The rotational relaxation time can accordingly be detected by dielectric relaxation spectroscopy but has also been obtained by NMR, enthalpy relaxation and other techniques. The relaxation times are strictly Arrhenius in character, as was also found recently[11] for the unusual case of the disordering dynamics in the binary alloy CoFe (1:1), which has a prominent "glass" transition at a temperature of $\sim 0.5 T_\lambda$[12] where frozen in disorder begins to equilibrate. Two further cases, more closely connected to the molecular rotators of Figure 1, are the simple branched hydrocarbon 1,4 dimethyl cyclohexane[13] and the group of ionic crystals represented by cesium nitrite[14], each of which exhibits a sharp lambda transition at temperatures far above the temperature of feeble manifestation of the glass transition. The first two of these thermodynamic manifestations of disordering phenomena are shown in Figure 2. The fact that each one exhibits[10, 11] the same Arrhenius kinetics of disordering manifested by the archetypal strong liquid, silica (in Figure 1), provides the stimulus for the following enquiry into the true nature of the disordering process in strong liquids, and its relation to the strong/fragile pattern of behavior in glassforming liquids (and plastic crystals) seen in Figure 1.

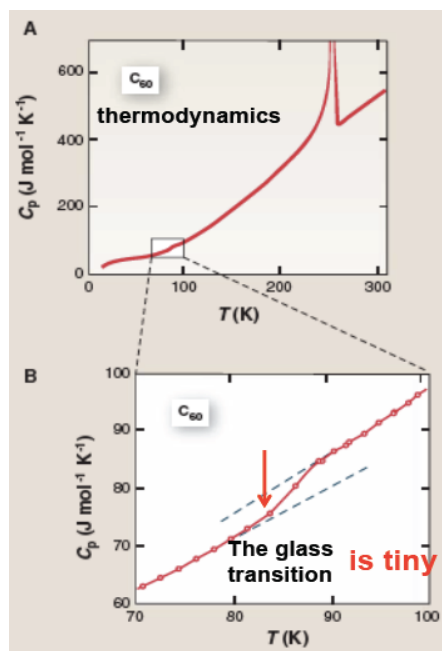

**FIGURE 2.** Two examples of crystalline systems that exhibit glass transitions as an obvious consequence of the arrest of a disordering process that culminates, at higher temperatures, in a cooperative (lambda) transition. (a) C 60

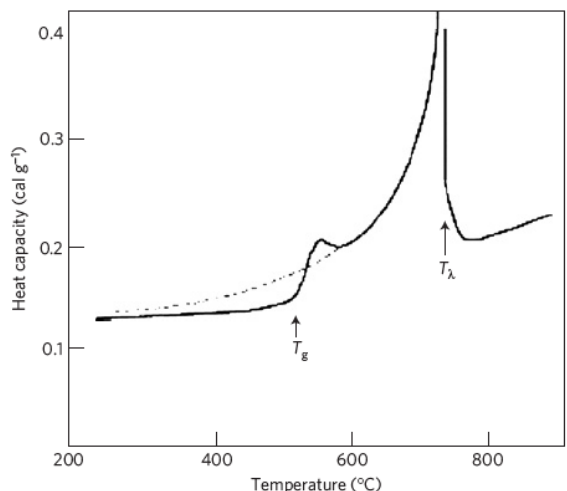

**Figure 2(b).** Co:Fe (1:1) adapted from ref.. (*5*). Glass transition is emphasized by upscanning much faster than cooling.

The first step is to enquire into the *thermodynamic* manifestation of condensed phase configurational excitation above $T_g$. There are insufficient data for a proper evaluation in the case of plastic crystals, although the heat capacity jumps are in the same order as the fragilities[6] (as is often, but certainly not always, observed for the case of liquids). For the liquids, there are data available to make the more correct comparisons based on comparison of the increasing entropy of the liquid state above $T_g$ (proportional to the jump in heat capacity) *after scaling by the excess entropy over crystal at the glass transition.* It is this scaling that was used[15] in comparing viscosity and entropic manifestations of fragility for a broad selection of glassforming liquids in an effort to understand the origin of the pattern of Figure 1, and it is this scaling that rectifies the inconsistencies pointed out by ourselves[16] and others[17] concerning the large heat capacity jumps of hydrogen bonding substances, despite their being not fragile, but only intermediate liquids. Of course, it turns out that these are also the cases with large excess entropies at $T_g$, so that the inconsistencies are removed in the scaled entropy plots.

Our scaling is to be rationalized by recognizing that in the two cases of volume-related response functions, (thermal expansivity and isothermal compressibility), the simple temperature or pressure coefficient for the response in question is scaled by the volume at each temperature (or pressure), so that for instance the expansivity, $\alpha = 1/V(\partial V/\partial T)_p$, and not simply $(\partial V/\partial T)_p$ like the $(\partial H/\partial T)_p$ that usually defines the heat capacity. Clearly $(\partial H/\partial T)_p$ should be normalized by some extensive property of the system but unfortunately there is no absolute value of the enthalpy by which to scale. However this problem is easily solved by defining the heat capacity as its entropy derivative, $C_p = (\partial S/\partial \ln T)_p$ (i.e. $[dH/T].[T/dT]$) - and then the appropriate comparison with expansivity and compressibility is $(1/S)(\partial S/\partial \ln T)_p$ where the extensive property again has an absolute value, thanks to the third law of thermodynamics. So if we want to make comparison of the way the excess entropy increases as T rises above $T_g$, then that entropy should be the integral of the S-scaled heat capacity, as was done[15].

Thus we emphasize, using the data of ref [15] that the strong/fragile pattern of behavior that we seek to elucidate is just as clearly manifested in the thermodynamic properties of the glassformer, as it is in the kinetics, (though the latter data are certainly easier to acquire). *Seeing the problem as a problem in thermodynamics, clears the way for us to seek some basis for the* *strong/fragile behavior of liquids in terms of thermodynamic transitions.*

## Is There a Thermodynamic Transition Underlying the Strong Liquid Dynamics and Thermodynamics?

In Figure 3 we present the available heat capacity data on silica, recognizing that the laboratory data above $T_g$ are seriously limited by the extremely high glass temperature (1470K for $SiO_2$) and by the difficulty of obtaining completely dry samples for study. Figure 3 takes advantage of the extensive and sophisticated molecular dynamics study, by Scheidler et al[18], of the model of silica known as "BKS" silica, after the partial ionic charges model of van Beest et al[19] that has proved so successful. This study provided both real and imaginary parts of the dynamic (complex) heat capacity which they combined with earlier laboratory and simulation data to yield Figure 3. We have included a well-defined point from their data at 6500K, and also include, as an insert, the corresponding behavior of the analog system $BeF_2$. Note that for $BeF_2$ there are more extensive laboratory data in the liquid state. The combination of data of Tamura[20, 21] and simulation data of ref.[22] has been seen before in ref[22]. Here it serves to confirm that the maximum in the heat capacity for $SiO_2$ is a real feature of the system thermodynamics.

The heat capacity maxima seen in these two systems of Figure 3 are certainly not lambda transitions, but certainly can be manifestations of lambda transitions that occur at higher pressures. Lambda transitions are critical points of the same universality class as the Ising model, and when they occur in systems with no crystal lattices to support them, they become liquid-liquid critical points. This phenomenology has been most intensively studied in the case of water, which is not a glassformer except under drastic quenching conditions - a consequence of the fact that the critical point occurs below the melting point and the large entropy fluctuations are very favorable for homogeneous nucleation.

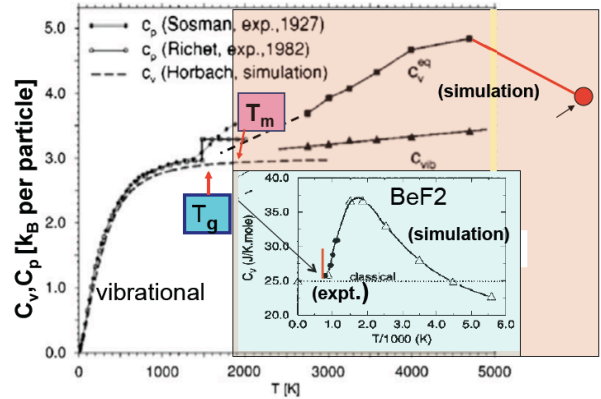

**Figure 3.** Thermodyamic features of the strong liquids $SiO_2$ and $BeF_2$ at temperatures far above the glass transition (and also the melting points) that we will relate to the lambda transitions associated with the strong plastic crystal and disordering alloy transitions. The main part of the diagram is from ref[18] and the additional point at high temperature, present in the raw data, is included to make the maximum present in the heat capacity function clear. This form is seen more clearly in the data for the analog (1/2 charges) $BeF_2$ model shown as an insert.

The possibility that a liquid-liquid critical point exists in $SiO_2$ at higher pressures has been pursued in earlier work[23] on the simple WAC potential[24] (only coulomb and exponential repulsion terms), with an interest in deciding whether or not the critical point hypothesized for water[25] might also be found in other tetrahedrally coordinated liquids (a more limited objective than the one we have embarked on here). The results obtained were very suggestive, but the computation times involved in equilibrating were too long to confirm that a critical point exists. More extensive studies by others[26, 27], using both WAC and BKS potentials, led to the conclusion, again by inference rather than direct observation, that there should be a critical point in each system, at 4000K for the WAC system and at a lower temperature, 2000K, for the BKS system.

In an ongoing study of the WAC system using a fast graphic card-based computer, we[28] have now observed the P vs. V isotherm for WAC $SiO_2$ as it flattens out (as required at a critical point) within the computational uncertainty, though more extensive computations are certainly needed to be fully convincing. The data acquired to date are shown in Figure 4. From the data seen in Figure 4, it is reasonable to assert that a flattening of the P-V isotherm, implying an infinity in the compressibility, has been observed at 3,500-4,000K. The pressure at which the maximum in the compressibility is found, is clearly a function of temperature, as was predicted in the metastable phase diagram deductions of Saika-Voivod et al[26] and the critical point is located at a pressure slightly lower than their estimate, viz. 5 GPA (vs 6.5 GPa).

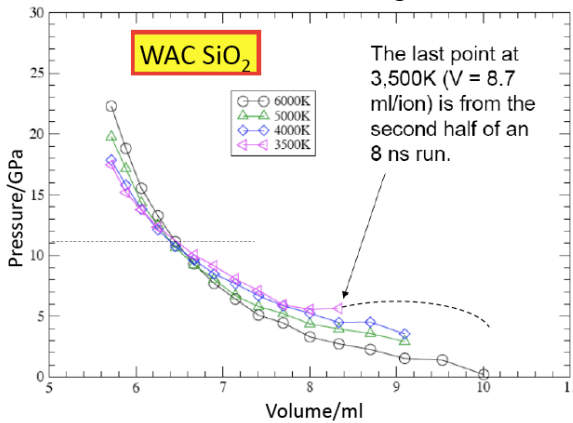

**Figure 4.** Preliminary data for the simplest silica model WAC extended down to 3500K to reveal the flattening of P vs V isotherm at 4000K and the onset of a van der Waals-like loop at lower temperatures.

The phase diagram from Saika-Voivod et al., with our compressibility maxima temperatures from Figure 4 (which are points on the Widom line extending beyond the coexistence line), is shown in Figure 5. Figure 5 also contains two panels showing the form of the heat capacity (obtained from the varying slope of the enthalpy vs temperature relation) both above (at 10 GPa) and below (at 0 pressure) the critical point pressure. The change in shape is quite striking, and it is completely consistent with the earlier demonstrations[27,29]) that the effect of pressure is to change, isothermally, the low temperature behavior of silica from that of a strong liquid to that of a fragile liquid. (Even ambient pressure silica becomes a fragile liquid on the *high* temperature side of the continuous and spread out. The drop in heat capacity of the fragile state at 10 GPa, unlike that of the low pressure liquid (with its F-S transition) is strictly due to ergodicity breaking. Its form is that seen for the excess $C_p$ of all *fragile* liquids[4].

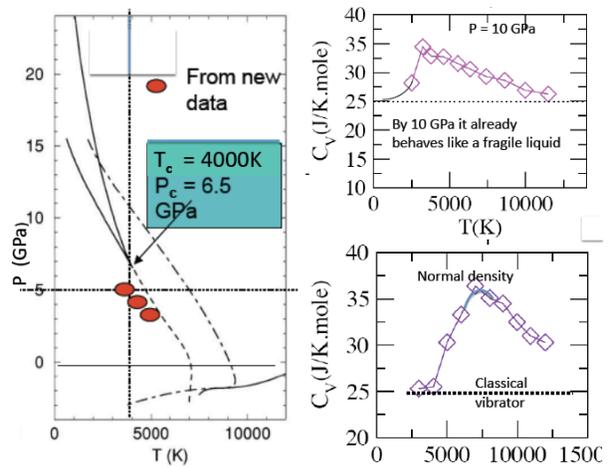

**Figure 5.** The phase diagram for liquid behavior in the system $SiO_2$ in the WAC potential, deduced by Saika-Voivod et al, to which are added the points of maximum compressibility according to Figure 4. These define the Widom line extension of the coexistence line and deviate slightly from the anticipated values but are in good qualitative agreement with them. The critical temp. Tc, (best estimate from Figure 4), is as predicted. Right hand panels show the heat capacity derived from the enthalpy vs. T isobars at pressures below and above the critical pressure. The zero pressure heat capacity has a rounded form as seen already in Figure 3, while the heat capacity at 10 GPa is sharply peaked and drops only as ergodicity is broken. This form is the well-known form for fragile liquids.

The behavior at the critical point cannot be reproduced by any computer simulation because the fluctuation lengthscales and timescales both diverge at the critical point. However, the form of the heat capacity, along the critical isobar, is required to be that of the transitions seen in Figure 2. An approach to this behavior for liquids has been seen recently in the detailed study of the attractive Jagla model[30] in which the second critical point can be parameterized to fall in the *stable* domain of the phase diagram[31].

This completes our discussion of the way in which the observations of Figure 1(b), relate to the behavior of ideally strong liquids. The conclusion is that an ideally strong liquid (of which glassy water is the best example though it can only be studied near its $T_g$) is one that in principle is approaching a lambda point which is in this case a liquid-liquid critical point. But there is an important additional implication of Figure 5 to be discussed. This relates to the metastable equilibrium behavior to be expected on the cooling the high pressure, fragile, liquid state of silica. It is as follows.

**LIQUID-LIQUID PHASE TRANSITION ON COOLING THE FRAGILE LIQUID STATE**

The phase diagram, Figure 5, that our present work semi-quantitatively supports, requires that on slow (equilibrated, isobaric) cooling over the range 5000K to 2000 K at 10 GPa pressure, the liquid should cross the coexistence line for the two forms of liquid silica of different density. The point at which the second liquid must form is seen more clearly in the final figure of this paper.

At that temperature, about 2,500K, there should be a first order phase change to a liquid of lower density (implying a volume expansion if the pressure is maintained constant). If the equilibrium phase change is inhibited by slow dynamics then the process should happen spinodally (no energy barrier) at the lower-lying spinodal limit to the mechanical stability of the high temperature liquid phase (seen in the Figure 5 phase diagram: the coexistence line between the two spinodals is not marked to avoid congestion[26]). Such a spinodal transformation to a phase of different density has been described in detail by Kurita and Tanaka[32] for the case of the fragile liquid, triphenyl phosphite. Indeed Tanaka[33] has given reasons for expecting it to be rather general in occurrence, (though occurring under unusual conditions).

If the transition does not occur at the coexistence line then the density temperature profile along this trajectory must have the form shown in Figure 6. Figure 6 is of course just the density vs. temperature equivalent of Figure 4. It shows a van der Waals type loop that has an unstable domain that, under equilibrium conditions (probably *metastable* with respect to some crystal, in the present case), would undergo the phase transition indicated by the vertical line.

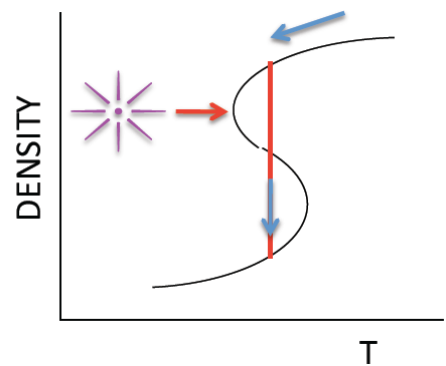

**Figure 6.** Density temperature relations during cooling at 10 GPa in Figure 5, showing the (metastable) equilibrium course involving a first order phase change with density decrease (vertical red line), and the spinodal terminus to supercooling, indicated by horizontal red arrow.

While the above might seem like pure speculation, it is to be pointed out that observations from the fibre optics community demonstrate that this temperature range (and in all probability also this pressure range), is actually encountered under conditions of a dramatic optical fiber failure phenomenon[34, 35], often called an "optical fuse"[36]. The failure[35, 37], which can destroy optical fibers by the kilometer, typically produces a moving plasma instability that is initiated when the transmitting light beam encounters a surface site that has been critically damaged[35, 36]. This is not a rare event in current high power applications of fiber optics. The plasma moves through the fibre towards the light source at about 1m/s, generating transiently huge temperature excursions in the fibre core. These must also be extremely high pressure excursions since the high strength fibre does not burst.

From the emission spectra of the plasma, it is deduced that the temperature reach >5,500K[37] with peaks to 10,000K, a condition that must generate a dense, probably supercritical, fluid phase, in which many valence states are explored. The uniaxial tensile strength of silica glass is well known to be in excess of 7GPa and the isotropic tensile limit seems also to be in this range, so the cooling plasma may well explore both of the critical points of the system (i.e. liquid vapor and liquid-liquid) critical conditions that then may ultimately be accessible to sophisticated investigation. As the plasma moves it must leave behind a rapidly cooled material, the structure and properties of which have not yet been given any attention by the polyamorphism community, though they certainly deserve it.

Our consideration of the phase relations for $SiO_2$ has so far been based on the behavior of the simple WAC model. It does not require much more CPU-time to study silica in the more realistic BKS model[19], in which the Si and O ionic charges are reduced and the cohesion decrease compensated by inclusion of a shorter range van der Waals interaction.

This case was also studied by Saika-Voivod et al.([26, 27]) and the phase diagram was found to be qualitatively similar to that of Figure 5, with a second critical point at comparable pressure but much lower temperature. When this point is placed on the known phase diagram for $SiO_2$ at high pressures, the second $T_c$ is located in the coesite domain of the phase diagram and well below the melting point. Not enough is known about the crystallization kinetics of silica at these temperatures and pressures to decide whether the critical point and coexistence line will be accessible experimentally (or preempted by crystallization as for water or hidden by glass transition, as discussed below). Hopefully, study of the fibres that have been subject to optical fusing might provide some answers.

**Extension to supercooling of fragile liquids of the normal type.**

The question now arises of the relation between the phenomena we have described above for the case of highly compressed liquid silica, a *fragile* liquid, and the behavior of the more common uncompressed molecular and ionic fragile liquids. There is, after all, a theoretical prediction that supercooling molecular liquids, of the highly fragile variety, will pass through the glass transition shortly before reaching a first order phase transition at which the cold liquid would, under (inaccessible) equilibrium conditions, pass to a glassy state of low excess enthalpy[4]. This is, of course, the *enthalpy* equivalent of the course of *density* change that we described above for silica (Figure 6). Let us remember that the density behavior of silica is anomalous, like that of water. The decrease in enthalpy in the case of ordinary van der Waals glassformers would be accompanied by an *increase* in density. To see how this would be simply related to the inversion of the Clapeyron slope for the liquid-liquid coexistence line, we compare the part of the diagram for silica containing the critical point and the coexistence domain, (Figure 5) with that anticipated for more normal liquids where a *positive* slope coexistence curve would be expected to exist. The figure suggests that the critical point, and its (Widom line) extension, could be encountered at high temperatures and pressures depending on the relative slopes of the glass temperature and the coexistence line. There is already evidence that the "glacial phase" of (normally very fragile) triphenyl phosphite can undergo a *spinodal* transformation to a more dense liquid phase under rapid cooling[32] and there is some suggestion that the "ultrastable glass" state (accessible by taking advantage of high surface mobility under favorable vapor deposit conditions) might be a further manifestation of a Figure 7 type phase behavior. This notion has recently been supported by observations on the reheating behavior of ultrastable toluene which seems to return to the normal viscous liquid state by a nucleation and growth kinetics[38] with Avrami coefficient of unity. Toluene is the case that, among common fragile liquids, tends most closely to encounter a coexistence line, or alternatively a spinodal collapse, according to the theoretical analysis of Matyushov and one of the present authors [4]

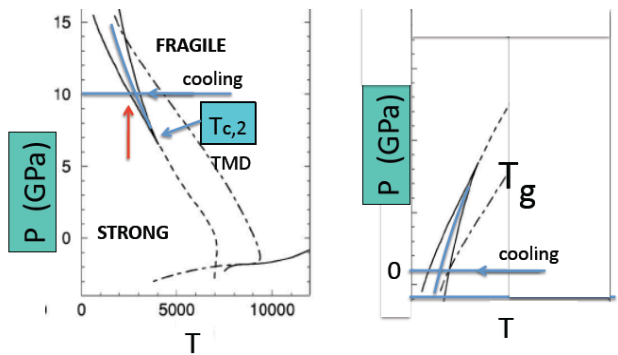

**Figure 7.** Comparison of the portion of the silica phase diagram where fragile liquid behavior is encountered (from Figure 5), (and a liquid-liquid transition should be encountered on sufficient cooling at 10 GPa pressure), with the equivalent diagram for liquids that have normal volume vs. enthalpy relations. A liquid-liquid transition is predicted below $T_g$, and a spinodal overshoot above $T_g$ on reheating, if the transition to the high density phase has been induced e.g. by vapor deposition near, but below, $T_g$.

## Conclusions

The interpretation of the "strong" behavior of the plastic crystal analogs of glassforming liquids in terms of the presence of an underlying order-disorder transition of the Ising universality class, offers a different approach to the interpretation of the behavior of the corresponding strong liquids. There is a strong implication that, without the constraints of a crystal lattice, the lambda transition of excitable crystal lattices can become the elusive second critical point of liquid state polyamorphism. Extensions to low temperatures of computer simulations of simple silica models produces evidence for the reality of such a second critical point in this archetypal strong liquid. The study of catastrophically damaged silica optical fibers may contain evidence for this critical complexity in laboratory liquid silica's behavior at high pressures. A simple inversion of the behavior discovered here for strong liquids offers an interesting possibility for connecting to theories for the behavior of fragile liquids near and below $T_g$, but much work needs to be done before such a possibility will be considered plausible.

## ACKNOWLEDGMENTS

This work was supported by the National Science Foundation experimental chemistry program under collaborative Grant no. CHE0404714. CAA acknowledges valuable monthly discussion with Pablo Debenedetti, Gene Stanley, and Peter Rossky. We are grateful to the participants in the "1st Advanced School on Materials for Photonic Applications" organized by Sidney Ribeiro in Araraquara, Brazil (October 2012), and particularly to Raman Kashyap (refs.[34, 35] for introducing us to the phenomenology of damage-induced plasmas in optical fibres, and for discussions of its possible value in elucidating some of the questions raised in this paper. Finally, we wish to acknowledge fruitful interactions with colleague Dmitry Matyushov, related to the fragile liquid sub-$T_g$ transition.